\documentclass[sigconf,nonacm]{acmart}  
\usepackage{threeparttable}
\usepackage{multirow}

\setcopyright{none}
\usepackage{url}
\usepackage{balance}
\begin{document}

\title{Fine-Tuning Multilingual Language Models for Code Review: An Empirical Study on Industrial C\# Projects}

\author{Igli Begolli}
\authornote{These authors contributed equally to this work.}
\affiliation{%
  \institution{Technical University Dortmund, Lovion GmbH}
  \city{Dortmund}
  \state{NRW}
  \country{Germany}}

\author{Meltem Aksoy}
\authornotemark[1]
\authornote{Corresponding author.}
\affiliation{%
  \institution{Research Center Trustworthy\\ Data Science and Security\\University Alliance Ruhr, Technical University Dortmund}
  \city{Dortmund} 
  \state{NRW}
  \country{Germany}}
  \email{meltem.aksoy@tu-dortmund.de}

\author{Daniel Neider}
\affiliation{%
  \institution{Research Center Trustworthy\\ Data Science and Security\\University Alliance Ruhr, Technical University Dortmund}
  \city{Dortmund} 
  \state{NRW}
  \country{Germany}}

\renewcommand{\shortauthors}{Begolli et al.}
\begin{abstract}
Code review is essential for maintaining software quality but often time-consuming and cognitively demanding, especially in industrial environments. Recent advancements in language models (LMs) have opened new avenues for automating code review tasks. This study presents the empirical evaluation of monolingual fine-tuning on the performance of open-source LMs across three key automated code review tasks: Code Change Quality Estimation, Review Comment Generation, and Code Refinement. We fine-tuned three distinct models—CodeReviewer, CodeLlama-7B, and DeepSeek-R1-Distill—on a C\#-specific dataset combining public benchmarks with industrial repositories. Our study investigates how different configurations of programming languages  and natural languages in the training data affect LM performance, particularly in comment generation.
Additionally, we benchmark the fine-tuned models against an automated static analysis tool (ASAT) and human reviewers to evaluate their practical utility in real-world settings. Our results show that monolingual fine-tuning improves model accuracy and relevance compared to multilingual baselines. While LMs can effectively support code review workflows, especially for routine or repetitive tasks, human reviewers remain superior in handling semantically complex or context-sensitive changes. Our findings highlight the importance of language alignment and task-specific adaptation in optimizing LMs for automated code review.
\end{abstract}

\maketitle
\section{Introduction}
Code review is a cornerstone of modern software engineering, serving multiple purposes including improving code quality, enforcing standards, detecting defects early, and promoting knowledge sharing across teams~\cite{tufano2021towards, fagan2002history, mantyla2008types}. This systematic practice has evolved into an integral part of development workflows, functioning both as a quality assurance mechanism and as a collaborative tool for continuous improvement.

Beyond technical benefits, code reviews also serve a social function. By enabling developers to inspect each other's changes, they facilitate informal learning, especially for junior team members, and help maintain consistency in project-specific practices~\cite{zhao2023survey}. Additionally, they contribute to software maintainability and reduce the risk of costly bugs~\cite{rahman, rigby, Mcintosh, bavota}.

Despite its numerous benefits, manual code review remains time-consuming, cognitively demanding, and difficult to scale effectively, particularly in large industrial projects~\cite{thongtanunam2022autotransform, pornprasit2023d}. Developers spend approximately 3–6 hours per week on code review tasks~\cite{cihan2024automated}. In large projects, reviewer assignment delays can postpone approvals by up to 12 days~\cite{thongtanunam2015}. The scale of this challenge is evident in major companies that process thousands of reviews monthly, with projects like Microsoft Bing handling approximately 3,000 reviews per month~\cite{rigby}. This demonstrates the substantial manual effort required and its potential to significantly impact development productivity.

Code review has undergone a significant transformation from traditional formal review approaches to today's collaborative, tool-supported methodologies~\cite{thongtanunam2015, abbood2020, cihan2024automated}. Automated static analysis tools (ASATs) are commonly deployed to reduce manual reviewing efforts by automatically detecting code smells, bugs, and coding standard violations. However, these tools frequently exhibit high false-positive rates and lack the contextual understanding required for nuanced code evaluations~\cite{vassallo2020developers, singh2017evaluating}. Consequently, developers must manually filter through numerous irrelevant warnings, which undermines tool effectiveness and user acceptance~\cite{li2022}. Current practices demonstrate significant effectiveness gaps, with only 15\% of review comments indicating actual defects and up to 34.50\% considered non-useful in major projects~\cite{li2022auger}.

Recent developments in artificial intelligence (AI), particularly in deep learning and natural language processing (NLP), have sparked increasing interest in automating the code review process, commonly referred to as automated code review (ACR). These efforts are often driven by the use of language models (LMs)\footnote{We use the term Language Model (LM) to refer to any model trained to understand or generate textual data. A Pretrained Language Model (PLM) is an LM that has undergone a general-purpose pretraining phase on large-scale unlabeled data before being fine-tuned for specific downstream tasks. A Large Language Model (LLM) refers to an LM with a high number of parameters (often in the billions), enabling it to handle a wide range of tasks—either zero-shot or with minimal fine-tuning. An LLM can also be considered a PLM, but not all PLMs qualify as LLMs. We refer to the plural forms as LMs, PLMs, and LLMs, respectively, throughout this paper.}, which are trained to understand or generate natural language (NL) text. Various pretrained language models (PLMs) have been proposed to support ACR tasks, including generating review comments, suggesting improvements, and transforming code into reviewer-approved versions~\cite{tufano2021towards, thongtanunam2022autotransform, zhao2023survey}. However, most of these efforts have focused on monolingual models trained on a small set of dominant programming languages (PLs), such as Java and Python~\cite{tufano2022using, li2022automating}, leaving their applicability to less represented languages like C\# relatively underexplored. This leaves a notable research gap regarding the effectiveness of PLMs and LLMs for industrial-strength languages like C\#, which, despite being widely used in enterprise software development, has received limited attention in prior ACR studies.

The emergence of large language models (LLMs) has further expanded the capabilities of ACR systems by enabling more context-aware and semantically rich interactions with code changes~\cite{lu2023llama, lin2024}. These models can significantly reduce reviewers’ manual workloads by automating repetitive tasks and identifying subtle issues that might otherwise go unnoticed. In some cases, organizations have already reported measurable improvements in review efficiency and developer satisfaction after integrating such models into their workflows~\cite{cihan2024automated}. Nonetheless, training LLMs for a specific language from scratch remains infeasible for most settings, as it requires large volumes of annotated data—which is costly and difficult to acquire for many PLs.

To address these challenges, some research has turned to multilingual PLMs pretrained on diverse PL corpora (codebases in multiple programming languages). These models have been explored in a range of software engineering tasks, including code summarization, search, and translation~\cite{ahmad2021unified, chen2022transferability}. However, they often demonstrate performance inconsistencies across languages—likely due to differences in syntax, idioms, and language-specific coding conventions, as well as the uneven representation of PLs in the pretraining datasets. This variability highlights the need for a language-aware adaptation, particularly for sensitive tasks like code review.

Further complicating this landscape, most prior evaluations have relied on open-source datasets from platforms such as GitHub~\cite{zhao2023survey, li2022auger}. While these datasets offer scalability and ease of access, they often capture limited contextual diversity and may not accurately reflect the complexity, review workflows, and coding standards found in industrial software development environments.

Taken together, these observations highlight several underexplored areas: the limited applicability of LMs to C\#, the lack of multi-task evaluations, inconsistent comparisons across model types, and the limited availability of studies based on real-world, industrial code review data.

To overcome these limitations, we adopt a new approach by fine-tuning existing multilingual LMs on monolingual C\# data. Specifically, we evaluate three open-source models with distinct pretraining objectives and architectural characteristics: CodeReviewer \cite{li2022automating}, a transformer-based encoder–decoder PLM designed specifically for review comment generation and pretrained on multilingual code corpora; CodeLlama-7B \cite{codellama2023}, a decoder-only multilingual LLM pretrained on a broad range of general-purpose code tasks; and DeepSeek-R1-Distill~\cite{guo2025deepseek}, a multilingual instruction-tuned LLM optimized for multi-domain applications.

To assess the effectiveness of these fine-tuned models in realistic review settings, we structure our evaluation around three core tasks commonly encountered in ACR workflows: (1) Code Change Quality Estimation \cite{hellendoorn2021towards, li2022automating}, which determines the necessity of human review for a given code change; (2) Review Comment Generation \cite{li2022auger,li2022automating,tufano2022using,lin2024,lin2024leveraging}, where NL feedback is generated to guide developers; and (3) Code Refinement \cite{li2023codeeditor,lin2023cct5,lin2023towards,pornprasit2023d,thongtanunam2022autotransform,tufano2019learning}, where suggested improvements are automatically applied to the codebase. Due to computational resource constraints, we limited fine-tuning of CodeLlama-7B and DeepSeek-R1-Distill-Llama-8B to a single task. We selected Review Comment Generation, given its linguistic complexity and high practical relevance in real-world code review workflows. For each task, we compare the performance of the fine-tuned models against their original (non-fine-tuned) baselines to evaluate the added value of task-specific adaptation.

To systematically explore model performance across these tasks, we pose the following research questions: \textbf{RQ1.} How does monolingual fine-tuning affect the performance of LMs in detecting whether a code change requires human review? \textbf{RQ2.} How does fine-tuning different types of LMs on different PL/NL combinations affect review comment generation? \textbf{RQ3.} What is the impact of fine-tuning on the code refinement capability of LMs, in terms of producing accurate and functionally equivalent code revisions? \textbf{RQ4.} How do fine-tuned LMs compare to an ASAT and human reviewers in identifying review-worthy code changes and generating feedback?

Our primary contributions are as follows:
\begin{itemize}
\item \textbf{Cross-Paradigm Evaluation Framework:} We systematically compare three open-source language models with distinct pretraining objectives—a review-specialized PLM (CodeReviewer), a code-pretrained LLM (CodeLlama-7B), and a general-purpose instruction-tuned LLM (DeepSeek-R1-Distill)—against both an industrial-strength ASAT (SonarQube) and human reviewers across all core ACR tasks.

\item \textbf{Language-Aligned Fine-Tuning Protocol:} We demonstrate that aligning both the programming language (monolingual C\#) and natural language (English-only vs. bilingual vs. multilingual) of training data significantly impacts review comment quality.

\item \textbf{Beyond Automated Metrics:} We complement standard BLEU-based evaluation with expert-validated human assessments (Information, Relevance, Issue Correctness Rate) on 40 production PRs, revealing that lexical similarity often misaligns with practical review utility. This dual-scope evaluation—combining large-scale automated metrics with human-aligned quality measures—addresses known limitations of surface-level NLP metrics in code review contexts.

\item \textbf{Industrial C\# Benchmark:} We introduce the first fine-tuned LM evaluation on enterprise-grade C\# repositories, addressing a critical gap in ACR research which has predominantly focused on Java and Python in open-source settings.
\end{itemize}
\section{Background and Related Work}
\subsection{Code Review Process}
Modern code review is a collaborative quality assurance practice where developers evaluate proposed code changes before integration into the main codebase~\cite{abbood2020,rahman}. Reviewers assess whether the submitted code satisfies both functional requirements (e.g., correct compilation and test coverage) and non-functional requirements such as readability, maintainability, and adherence to coding conventions.

The process typically involves analyzing source code written in a programming language (PL) and formulating feedback in natural language (NL). On platforms such as GitHub, Gerrit, and Phabricator, the review workflow follows five main steps. First, a contributor submits a patch through a pull request (PR). Then, one or more reviewers, ideally with relevant domain knowledge, examine the code diff and provide NL comments, along with approval or rejection votes. Based on this feedback, the contributor modifies the code. This review cycle iterates until the code meets predefined quality standards or the submission is discarded.
\subsection{Automated Static Analysis Tools (ASATs) for Code Review}
Manual code review remains a fundamental yet resource-intensive practice in software development. To reduce the manual effort required in this iterative process, a variety of automated approaches have been developed, with ASATs being among the earliest and most widely adopted solutions. ASATs enable developers to identify potential code issues—such as bugs, syntax violations, and deviations from best practices—without executing the code \cite{ernst2015measure}. Widely used tools like SonarQube \cite{sonarqube} and PMD \cite{pmd} employ rule-based mechanisms to flag such issues early in the software development lifecycle, thereby supporting both quality assurance and coding standard enforcement.

\citet{vassallo2020developers} emphasize that successful integration of ASATs into developer workflows and trust in their outputs are key factors affecting their practical utility. \citet{beller2016analyzing} further observe that tool performance can vary depending on the PL, highlighting the role of contextual factors in tool effectiveness.

ASATs are particularly useful in identifying superficial defects such as style inconsistencies and common programming errors \cite{panichella2015would,mehrpour2023can,singh2017evaluating}. As reported by \citet{singh2017evaluating} report ASATs can automatically detect up to 16\% of issues later identified by human reviewers, suggesting their potential to reduce reviewer workload. However, ASATs often struggle with more complex concerns—such as architectural or domain-specific flaws—and suffer from high false-positive rates that may lead to reviewer fatigue and declining trust \cite{charoenwet2024empirical}.

While ASATs offer valuable support in code review by automating the detection of routine issues, their inability to address nuanced, context-dependent problems highlights the need for complementary solutions. To this end, our comparative evaluation includes SonarQube as a representative ASAT, selected for its strong performance baseline and widespread adoption in industry \cite{lenarduzzi2023critical}. These limitations further motivate the investigation of more advanced approaches capable of handling the contextual and semantic complexity inherent in real-world code review scenarios.
\subsection{Automated Code Review (ACR)}
While ASATs help detect rule-based issues like style violations, simple bugs, and security vulnerabilities, they fall short in providing context-aware, semantic, or design-level feedback \cite{tufano2021towards, tufano2024code, hellendoorn2021towards}. To address these limitations and reduce developers' cognitive load, the software engineering community has increasingly explored AI-driven and NLP-based solutions.

Early ACR research primarily focused on isolated quality aspects, including bug detection \cite{mantyla2008types}, vulnerability identification \cite{bosu2014identifying, thompson2017large}, and style inconsistency detection \cite{paixao2020behind}, demonstrating positive effects on software maintainability \cite{panichella2015would, beller2016analyzing}.

Building on these foundations, recent work in ACR has focused on three core tasks that collectively reflect the broader goals of the code review process.

The first is \textit{Code Change Quality Estimation}, which aims to predict whether a given code change requires human review. Initial approaches relied on handcrafted features and shallow learning models \cite{Jiang}, whereas more recent studies have employed deep learning classifiers trained on semantic representations of code diffs to improve prediction accuracy \cite{Li2019, hellendoorn2021towards, li2022automating}.

The second is \textit{Review Comment Generation}, widely considered the most linguistically and semantically demanding aspect of ACR. Unlike tasks such as code refinement, this task requires generating context-aware, human-like feedback based on limited code context. Task-specific PLMs, such as T5-Review \cite{tufano2022using}, built on the T5 architecture \cite{raffel2020} and trained on datasets like Stack Overflow and CodeSearchNet, often struggled to match the performance of refinement-focused models in generating high-quality comments. To improve the semantic alignment between code functionality and review comments, AUGER \cite{li2022auger} introduced a joint modeling strategy that links code functionality with relevant review comments, leveraging pretraining techniques such as denoising and comment summarization to improve feedback relevance. Retrieval-based models like CommentFinder \cite{hong2022} further offered efficient, non-generative alternatives by surfacing relevant comments from historical data.

The third is \textit{Code Refinement}, which focuses on automatically generating code changes in response to reviewer feedback. Models such as Trans-Review \cite{tufano2021towards} applied sequence-to-sequence learning with source code abstraction \cite{tufano2019learning} to reduce vocabulary size. AutoTransform \cite{thongtanunam2022autotransform} improved identifier representation using Byte-Pair Encoding (BPE, a token compression technique) \cite{sennrich2015}. T5-Review \cite{tufano2022using} enhanced performance by leveraging large-scale code-text pretraining. Later methods, including CodeEditor \cite{li2023codeeditor} and D-ACT \cite{pornprasit2023d}, focused on learning from code edits and diff-awareness.

Moving beyond isolated task formulations, recent research has explored multi-task learning and large-scale pretraining to support the entire code review pipeline. For instance, CCT5 \cite{lin2023cct5} was trained on 1.5 million diff–comment pairs and designed to address both review comment generation and code refinement. \citet{lin2024} proposed experience-aware oversampling to emphasize high-quality human reviews, improving model performance across multiple review stages. A large-scale benchmark by \citet{zhao2023survey} systematically compared three task-specific PLMs (Trans-Review, AutoTransform, and T5-Review) with general-purpose code PLMs (CodeBERT \cite{feng2020codebert}, CodeT5 \cite{codet5}) across all ACR tasks. Their findings showed that CodeT5 achieved the highest performance in code refinement, whereas T5-Review outperformed others in review comment generation, highlighting the advantages of task specialization for linguistically complex review tasks.

The emergence of LLMs has further broadened the scope of ACR. \citet{lu2023llama} presented LLaMA-Reviewer, which leverages LoRA (Low-Rank Adaptation), a parameter-efficient fine-tuning method that updates only a subset of model weights, to support all three ACR stages without full retraining. \citet{guo2024} found that ChatGPT, despite lacking task-specific fine-tuning, outperformed CodeReviewer in refinement tasks but underperformed in comment generation. Complementing these efforts, \citet{cihan2024automated} conducted one of the first in-situ evaluations of LLM-assisted code review in industry, reporting better comment resolution rates but mixed developer satisfaction and increased PR closure times.

Despite these advances, open-source LLMs continue to lag behind proprietary models trained on expert-curated industrial datasets \cite{lin2023cct5, vijayvergiya}. Moreover, comprehensive evaluations covering all three ACR stages remain limited—particularly in realistic, industry-grade C\# codebases and in direct comparisons with both ASAT and human reviewers. Recent works \cite{lu2023llama, guo2024} have started to examine the potential of general-purpose, instruction-tuned LLMs (e.g., ChatGPT, DeepSeek-R1-Distill) for code review tasks. Instruction-tuned models are trained to interpret and follow NL instructions, enabling them to generate appropriate responses to user prompts without requiring task-specific fine-tuning. These models show promising performance in reasoning and comment relevance, despite lacking explicit code-specific pretraining. However, direct comparisons between code-pretrained and general-purpose LLMs in industrial C\# settings remain scarce—highlighting a critical gap that this study aims to address. C\# remains a widely used language in enterprise software development, particularly in sectors such as finance, energy, and manufacturing~\cite{stackoverflow2023}. Despite its industry relevance, it remains underrepresented in academic datasets and ACR studies.

\section{Study Design and Methodology}
\begin{figure*}[ht]
  \centering
  \includegraphics[width=0.8\linewidth, height=0.5\textheight, keepaspectratio]{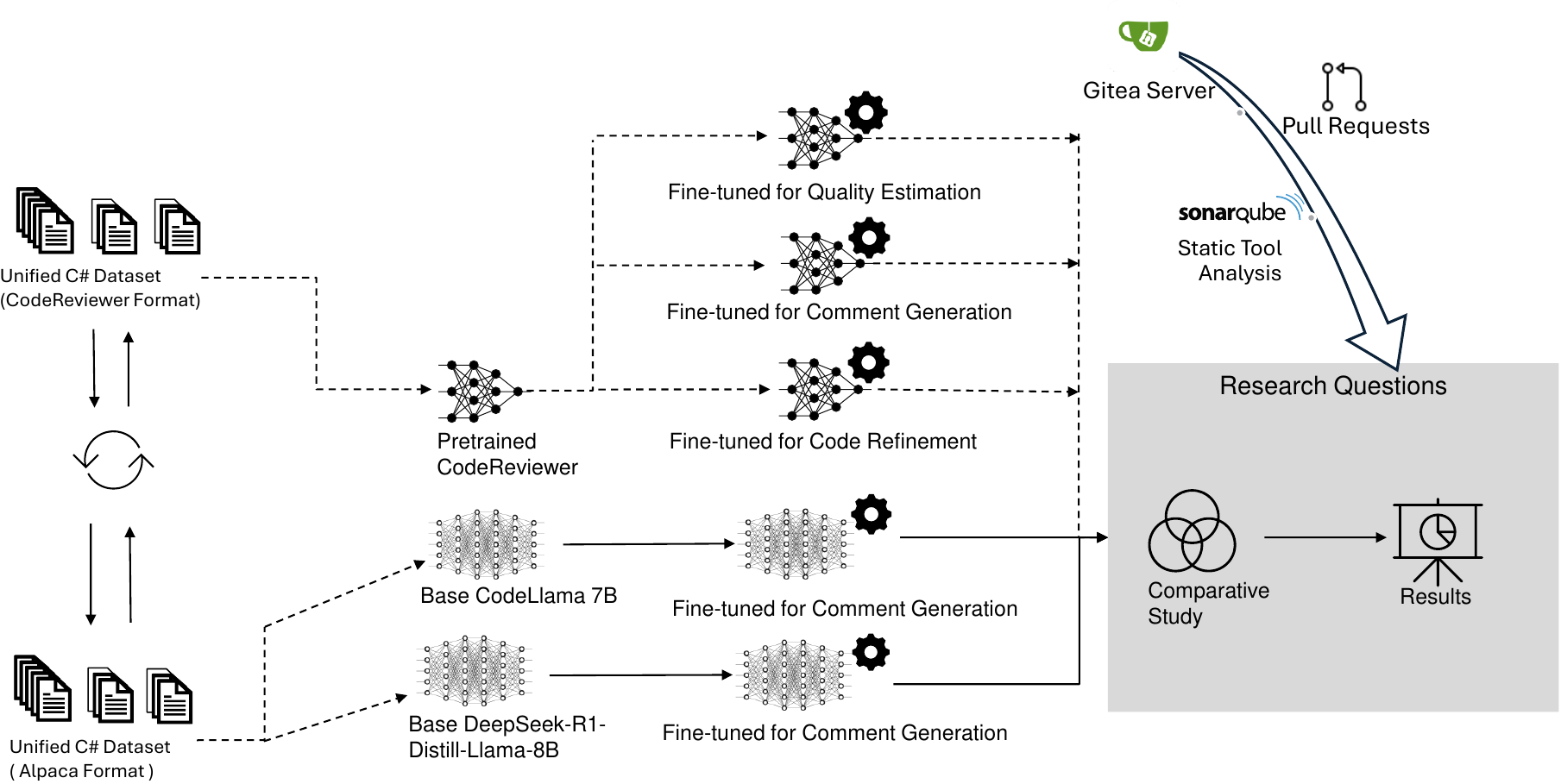}
  \Description{Overview of the experimental workflow.}
  \caption{Overview of the experimental workflow.}

  \label{fig:summary}
\end{figure*}

This section provides an overview of our experimental setup, including the environment, dataset construction, model fine-tuning, and evaluation framework. Figure~\ref{fig:summary} illustrates the overall workflow of our study.
\subsection{Study Context}
This study investigates the effectiveness of fine-tuning three different types of LMs on monolingual C\# data across three code review tasks in an industrial setting. Our research is motivated by the performance gap observed between public benchmark results and real-world industrial deployments of LM-based ACR systems \cite{vijayvergiya, frommgen2024resolving}. 

We collaborated with Lovion GmbH, a software company specializing in end-to-end digital solutions for infrastructure asset and network management. Their development teams primarily work with C\#, follow an agile methodology, and actively adopt the latest advancements in software engineering technologies.

Access to Lovion’s internal repositories provided us with high-quality, production-grade review comments written by experienced engineers. This unique dataset allows us to evaluate LLM performance under realistic conditions and explore how training data language composition (English-only vs. multilingual vs. translated) affects review comment generation quality.

By focusing on C\#, a PL with high industrial relevance \cite{rempel2016probabilistic}, we aim to provide practical insights for both researchers and practitioners on adapting LMs for industrial ACR tasks.
\subsection{Data Preparation}
\label{sec:data-preparation}
We built task-specific datasets by combining PRs and reviewer comments from five internal C\# repositories at Lovion GmbH. Data extraction from Lovion's repositories was conducted via Gitea’s REST API, ensuring comprehensive metadata coverage, including PR information, code diffs, and review comments.

These five repositories correspond to distinct modules of Lovion GmbH’s enterprise asset management platform, covering user interface, data access, visualization, workflow automation, and testing components. 
They differ in functionality and code complexity, providing a representative view of industrial C\# development practices. 
All repositories are actively maintained under Lovion’s internal quality assurance policies, including mandatory peer review and continuous integration (CI) checks, ensuring that the extracted PRs and comments are high-quality and production-grade.

To complement the industrial dataset, we integrated the publicly available CodeReviewer benchmark~\cite{li2022automating}. 
This addition enhanced data volume and supported stable model training. 
While the Lovion repositories provided rich, domain-specific review data representative of real industrial practices, their limited scale posed challenges for fine-tuning large transformer-based models, such as overfitting and unstable convergence. 
Incorporating the C\# subset of the CodeReviewer benchmark mitigated these issues by supplying additional language-consistent samples without introducing domain overlap. 
To prevent data leakage, we included only C\# examples and verified that both datasets originated from distinct platforms (Gitea vs. GitHub) with no shared PRs. 
All data were re-indexed, deduplicated, and filtered to remove incomplete or non-compilable code snippets before unification.

\textbf{Translation and Quality Validation.} Since the Lovion review comments were originally written in German, they were automatically translated into English using Python’s \texttt{googletrans} library (v4.0.0rc1, free API mode). To ensure translation fidelity, two bilingual annotators with professional English–German proficiency manually reviewed a stratified random sample of 200 translated comment pairs (\~6\% of the dataset). 
Each pair was rated on a 1–5 adequacy and fluency scale, yielding mean scores of 4.6 and 4.8, respectively. 
Inter-annotator agreement was substantial (Cohen’s $\kappa$ = 0.82), indicating consistent judgments. 
Disagreements were resolved through discussion until consensus was reached. 
Minor domain-specific inconsistencies (e.g., technical terms and C\#-specific vocabulary) were corrected during preprocessing, ensuring a high overall translation quality.

\textbf{Language Configurations.} 
The resulting dataset supports three NL configurations used across experiments: 
(i) \textit{English-only}, consisting of translated Lovion comments and English samples from the CodeReviewer dataset; 
(ii) \textit{English+German}, preserving both original and translated Lovion comments; 
and (iii) \textit{Multilingual}, incorporating additional non-English samples (e.g., Chinese, Spanish, and French) from the CodeReviewer benchmark. 
Irrelevant or low-quality instances were removed during preprocessing to ensure consistency across configurations. 
These configurations are later used in the Review Comment Generation task to assess the effect of language composition on model performance.

\textbf{Data Splitting and Formatting.} For each of the three downstream tasks, we created dedicated datasets. 
All PRs were first chronologically sorted by submission date. 
We then partitioned the data sequentially into training (85\%), validation (7.5\%), and test (7.5\%) subsets, ensuring that newer PRs were never included in the training data relative to the test set. 
Within each subset, samples were randomly shuffled to mitigate potential ordering bias.

Finally, we applied task-specific data formatting.
For the Code Change Quality Estimation task, we constructed a total of 44,962 samples, equally divided between positive (\(y=1\)) and negative (\(y=0\)) instances. 
Positive samples (\(y=1\)) were directly sourced from the Comment Generation dataset, corresponding to diff hunks that received human review comments. 
Negative samples (\(y=0\)) were drawn from the final PR versions where no reviewer comments were issued, representing either accepted or non-critical changes. 
This balancing procedure helped mitigate bias toward the majority class and ensured comparable representation of both categories. 
For Comment Generation, each instance included a code diff and its corresponding human-written review comment. For Code Refinement, each data point consisted of a before-and-after code pair reflecting reviewer-suggested changes.

Detailed dataset statistics across all tasks and sources are summarized in Table~\ref{tab:dataset-stats}.

\begin{table}[!htbp]
  \centering
  \caption{Dataset statistics across three tasks and data sources.}
  \label{tab:dataset-stats}
  \small 
  \setlength{\tabcolsep}{3pt} 
  \resizebox{\columnwidth}{!}{
  \begin{tabular}{llccc}
    \toprule
    \textbf{Task} & \textbf{Dataset} & \textbf{Train} & \textbf{Val} & \textbf{Test} \\
    \midrule
    \multirow{3}{*}{\shortstack{Code Change \\ Quality Estimation}}
      & Lovion & ~19,110 & ~1,685 & ~1,686 \\
      & CodeReviewer & ~19,110 & ~1,685 & ~1,686 \\
      & Unified & ~38,220 & ~3,370 & ~3,372 \\
    \midrule
    \multirow{3}{*}{Comment Generation} 
      & Lovion & ~3,420 & ~302 & ~302 \\
      & CodeReviewer & ~15,689 & ~1,383 & ~1,384 \\
      & Unified & ~19,110 & ~1,685 & ~1,686 \\
    \midrule
    \multirow{3}{*}{Code Refinement} 
      & Lovion & ~2,125 & ~187 & ~188 \\
      & CodeReviewer & ~13,848 & ~1,222 & ~1,222 \\
      & Unified & ~15,973 & ~1,409 & ~1,410 \\
    \bottomrule
  \end{tabular}}
\end{table}
\subsection{Experimental Setup}
We conducted all experiments on a GPU-accelerated server running Ubuntu 20.04, equipped with Python 3.12 and an NVIDIA A100 GPU (80 GB VRAM). For model development and fine-tuning, we used PyTorch 2.0 (CUDA 11.8) in combination with Hugging Face Transformers (v4.48). We implemented QLoRA (Quantized Low-Rank Adaptation) fine-tuning using the Axolotl and Unsloth frameworks. QLoRA enables efficient fine-tuning of large language models by combining parameter-efficient low-rank updates with quantization techniques that reduce memory usage. Additionally, we used NumPy and scikit-learn for data preprocessing and evaluation.
\subsection{Model Fine-tuning}
\label{sec:model-fine-tuning}
We fine-tuned three distinct models in this study: \textit{CodeReviewer}~\cite{li2022automating}, \textit{CodeLlama-7B}~\cite{roziere2023code}, and \textit{DeepSeek-R1-Distill-Llama-8B}~\cite{guo2025deepseek}.
\textit{CodeReviewer} is a transformer-based encoder-decoder PLM pretrained on multilingual code and review comment pairs. It was specifically developed for ACR tasks and thus represents a review-specialized PLM. \textit{CodeLlama-7B} is a decoder-only LLM pretrained on a wide range of code corpora in multiple PLs. It serves as a code-pretrained LLM optimized for general code understanding and generation. \textit{DeepSeek-R1-Distill-Llama-8B} is an instruction-tuned general-purpose LLM, distilled from LLaMA 3.1–8B, with no code-specific pretraining. It is designed to perform well across diverse NL tasks, including reasoning and instruction following. This diversity in pretraining paradigms enables a comprehensive comparison of review-specific, code-oriented, and general-purpose LMs in monolingual C\# review settings.

The selection of \textit{CodeReviewer}, \textit{CodeLlama-7B}, and \textit{DeepSeek-R1-Distill-Llama-8B} was guided by three main criteria: (i) \textit{open-source accessibility} ensuring full reproducibility and parameter-efficient fine-tuning; (ii) \textit{architectural diversity}, covering a review-specific PLM (CodeReviewer), a code-pretrained LLM (CodeLlama), and a general-purpose instruction-tuned LLM (DeepSeek); and (iii) \textit{representativeness} of SOTA open models widely adopted in automated code review research. Proprietary models such as GPT, Claude, or Gemini were not included due to closed-weight architectures, licensing restrictions, limited reproducibility in fine-tuning workflows and privacy concerns associated with uploading industrial code data to external APIs. Our focus is therefore on open, reproducible, and practically deployable models suitable for industrial integration.

\subsubsection{Fine-Tuning CodeReviewer for Three ACR Tasks}
We fine-tuned \textit{CodeReviewer} for all three ACR tasks based on its original multi-task design and prior performance in both classification and generation tasks~\cite{li2022automating}. We followed the original implementation as a basis and adapted task-specific hyperparameters where needed.
\begin{enumerate}
  \item \emph{Code Change Quality Estimation:} We fine-tuned CodeReviewer as a binary classifier to predict whether a code change requires a review comment (\( y = 1 \)) or not (\( y = 0 \)). Using a learning rate of \( 3 \times 10^{-4} \), we trained the model for 5 epochs with a batch size of 12. This configuration helped accelerate training while maintaining stability.
  \item \emph{Review Comment Generation:} We applied a two-stage fine-tuning approach. In the first stage, we used a mixed-language dataset that included German comments from the Lovion dataset and English comments from the CodeReviewer benchmark. In the second stage, we used the translated version of the Lovion comments (see Section~\ref{sec:data-preparation}) to retrain the model on a fully English dataset. Across both stages, the model was trained for approximately 3 epochs (around 7,500 steps) using a learning rate of \(3 \times 10^{-4}\) and a batch size of 6.
  \item \emph{Code Refinement:} For this task, we trained the model to generate improved code versions based on reviewer feedback. Fine-tuning was performed for 3–4 epochs using a batch size of 8 and a learning rate of \(3 \times 10^{-4}\), striking a balance between efficiency and convergence.
\end{enumerate}
\subsubsection{Fine-Tuning CodeLlama and DeepSeek for Review Comment Generation} 
In addition to the review-specialized \textit{CodeReviewer} model, we fine-tuned both \textit{CodeLlama-7B} and \textit{DeepSeek-R1-Distill-Llama-8B} specifically for the Review Comment Generation task.  Their instruction-tuned architectures and strong NL generation capabilities make them particularly well-suited for review comment generation. Due to computational constraints, we focused their evaluation exclusively on review comment generation to assess their ability to generate high-quality, context-aware review comments for C\#.

To enable instruction-based fine-tuning, we reformatted the dataset using the Stanford Alpaca prompt format~\cite{taori2023stanford}, in line with best practices for instruction-tuned LLM training~\cite{lu2023llama}. Each prompt included (i) an \texttt{instruction} field describing the task (e.g., ``Generate a review comment for the following code snippet''), (ii) an \texttt{input} field containing the code diff, and (iii) an \texttt{output} field with the corresponding human-written review comment. The full prompt template is provided in Table~\ref{tab:prompt-format}.
\begin{table}[!htbp]
\centering
\caption{Instruction-based prompt format used for fine-tuning.}
\label{tab:prompt-format}
\begin{tabular}{p{0.95\linewidth}} 
\hline
\textbf{Prompt Template} : Below is an instruction that describes a task, paired with an input that provides further context. Write a response that appropriately completes the request. \\
\#\#\#\textbf{Instruction}: \{instruction\}\\
\#\#\#\textbf{Input}: \{input\}\\
\#\#\#\textbf{Response}:\{output\}\\
\hline
\textbf{Instruction}\\ You are a powerful code reviewer model for the C\#. Your job is to suggest review comment in natural language. You are given a context regarding a diff hunk or code change in programming language. You must output appropriate, contextual review comment for that code change. \\
\textbf{Input}: Diff Hunk: \{diff hunk\}\\
\textbf{Output}: \{review comment\} \\
\hline
\end{tabular}
\end{table}

We fine-tuned CodeLlama and DeepSeek using 4-bit QLoRA, a technique that enables efficient fine-tuning of large-scale models by combining weight quantization with low-rank adaptation. The shared hyperparameters included 3 training epochs, a LoRA rank of 32 (i.e., the dimensionality of the trainable low-rank matrices), a token limit of 2,048, a dropout rate of 0.05, and a learning rate of 0.0002, optimized using the paged AdamW algorithm. For \textit{CodeLlama-7B}, we employed the Axolotl framework with a weight decay of 0.0, whereas for \textit{DeepSeek-R1-Distill-Llama-8B}, we used the Unsloth framework with a weight decay of 0.01.

\subsubsection{Training Data Scope and Language Composition.} 
Since both the programming language (PL) used in the code and the natural language (NL) used in review comments vary across models and datasets, we classified the training configurations along two dimensions:
\begin{itemize}
    \item \textbf{PL Scope:}
    \begin{itemize}
        \item \textbf{Mono-PL:} Only \texttt{C\#} code.
        \item \textbf{Multi-PL:} Multiple PLs (e.g., Java, Python, C\#) from the multilingual benchmark~\cite{li2022automating, haider2024prompting}.
    \end{itemize}
    \item \textbf{NL of Comments:}
    \begin{itemize}
        \item \textbf{English-only:} All comments translated and unified to English.
        \item \textbf{English+German:} Mixed comments in both languages.
        \item \textbf{Multilingual:} Diverse NL from the original multilingual benchmark.
    \end{itemize}
\end{itemize}
\subsection{Data and Code Availability}
All scripts and fine-tuning configurations used in this study are available in the accompanying replication package.\footnote{\url{https://anonymous.4open.science/r/CodeReview-LMs/}}
The package contains preprocessing scripts, evaluation code, and instructions to reproduce all results reported in this paper. 
The industrial C\# PR dataset from Lovion GmbH, however, cannot be publicly released due to confidentiality agreements with the company. 
To ensure transparency, we provide detailed metadata and aggregated statistics describing the dataset’s structure (e.g., issue types, PR size distribution, and module coverage) so that the experimental design and data characteristics can be replicated without accessing the raw code. 
Researchers interested in collaboration or controlled data access may contact the authors subject to company approval.

\subsection{SonarQube Integration}
SonarQube~\cite{sonarqube} is a widely adopted open-source ASAT that detects code quality issues across three primary dimensions: reliability, maintainability, and security. It identifies a variety of issues, including bugs, code smells, and security vulnerabilities. While SonarQube’s definition of code smells partially overlaps with the classic taxonomy introduced by \citet{fowler2018refactoring}, it also incorporates platform-specific classifications and rule extensions.

For this study, we configured SonarQube specifically for C\# by activating approximately 450 language-specific rules from the official SonarC\# rule set\footnote{\url{https://rules.sonarsource.com/csharp/}}. This ensured adherence to recognized best practices for C\# development, covering security, reliability, and maintainability.

While SonarQube was not designed to generate NL feedback, we included it as a complementary baseline to represent rule-based, non-linguistic analysis within our unified evaluation framework. This allowed us to systematically compare traditional static analysis with language-based review approaches in identifying review-relevant issues. To ensure comparability, we applied the same correctness criterion (i.e., “good” vs. “not good”) when evaluating its detections against ground truth. Hence, SonarQube serves not as a linguistic peer to LMs, but as a representative of established industrial practices for rule-based review support.

To integrate SonarQube into Lovion GmbH’s DevOps pipeline, we deployed the system within the company’s Jenkins-based continuous integration (CI) workflow. This setup enabled automated analysis of all relevant PRs and ensured consistent quality assessment outputs. These outputs served as a static baseline against which we compared both LM-generated and human-written code review results.

\subsection{Evaluation Framework}
\label{sec:evaluation-framework}
To comprehensively address our research questions, we adopt a dual-scope evaluation:  
(i) a \textit{full test set} for large-scale, fully automated metrics, and  
(ii) a \textit{40-PR annotated subset} for expert-validated, human-aligned analyses.  
This design provides both breadth (statistical comparability on all PRs) and depth (contextual quality against expert ground truth).  
We evaluate both fine-tuned and non–fine-tuned versions of each model.

\subsubsection{Full test set (automated metrics)}  
PRs from real workflows with existing human comments were used to compute automated metrics across all PRs:

- For \textit{Code Change Quality Estimation}, we used standard classification metrics—accuracy, precision, recall, and F1 score—to evaluate how accurately each model identified whether a PR required a review comment (\textit{problematic}) or not (\textit{acceptable}).

- For \textit{Review Comment Generation}, we used BLEU-4~\cite{papineni2002bleu} to measure lexical overlap between generated and reference comments.  
This choice aligns with the evaluation protocol in the \textit{CodeReviewer} study~\cite{li2022automating}, ensuring methodological comparability.  
Following prior works~\cite{li2022automating, codebert_smooth_bleu}, BLEU-4 scores were computed over the first 256 output tokens to standardize sequence length and reduce the bias introduced by lengthy generations.  
For \textit{DeepSeek-R1-Distill-Llama-8B}, chain-of-thought (CoT) reasoning segments were excluded prior to scoring for consistency.  
Although code-sensitive metrics such as CodeBLEU~\cite{ren2020codebleu} or CrystalBLEU~\cite{CrystalBLEU2023} could offer complementary insights, they were not used because the original \textit{CodeReviewer} checkpoints are not publicly available, making cross-study comparison infeasible.

- For \textit{Code Refinement}, we used BLEU to assess n-gram similarity and Exact Match (EM) to verify byte-level equivalence between generated and reference code revisions.  
These metrics were selected to remain consistent with the \textit{CodeReviewer} baseline setup and to enable one-to-one comparison of C\#-specific results.

\subsubsection{40-PR annotated subset (human-aligned metrics)}  
We selected 40 PRs from Lovion GmbH’s industrial C\# repositories using stratified random sampling to ensure balanced coverage across modules and issue types.  
Each PR contained a single code hunk to maintain a clear mapping between code change and review feedback.  
The subset consisted of 20 PRs that required review comments—covering issues categorized as \textit{bug}, \textit{performance}, \textit{maintainability}, or \textit{security}—and another 20 PRs that did not require any review comments, serving as a control group. Two senior engineers independently determined whether a review comment was required and assigned the appropriate issue category. Disagreements were resolved through discussion (Cohen’s $\kappa$ = 0.79). These expert-validated annotations established the \textit{ground truth} foundation for all subsequent model and human performance evaluations.

Distinct evaluation protocols were applied for each ACR task:

- For \textit{Code Change Quality Estimation}, LM and SonarQube predictions were compared against the expert-validated ground truth labels using standard classification metrics—accuracy, precision, recall, and F1 score.  
Human reviewers—six bilingual software engineers from Lovion GmbH (two senior, two mid-level, and two junior)—independently evaluated all 40 PRs to determine whether a review comment was required, and their majority decision served as the reference judgment for human performance.

- For \textit{Review Comment Generation}, we used the 20 PRs that required comments.  
Two senior engineers (the same annotators who established the ground truth) evaluated whether each system’s output—either a natural-language comment (for humans and LMs) or a detected issue (for SonarQube)—correctly addressed the issue category defined in the ground truth.  
Each output was labeled as \textit{“correct”} if it matched the issue type (e.g., bug, performance, maintainability, or security) and as \textit{“incorrect”} otherwise.  
The proportion of correct outputs constituted the \textit{Issue Correctness Rate}, providing a unified measure of issue-level correctness across rule-based and language-based systems.

The six human reviewers also provided qualitative evaluations of AI-generated comments for the same 20 PRs. They rated each comment on two 5-point Likert scales:  
\textit{Information} (extent to which the comment provides meaningful and actionable feedback) and \textit{Relevance} (degree to which the comment addresses the actual issue in the code diff).  
The final score for each comment was calculated by averaging the six individual ratings.  
To avoid potential bias, the reviewers were independent of dataset curation and ground-truth annotation, and the two activities—manual code reviewing and evaluation of AI-generated outputs—were conducted separately to prevent cross-contamination of judgments.  
The aggregated human ratings thus represent a consensus-based human baseline that serves as the reference point across all evaluation dimensions.

\subsubsection{Efficiency}  
Efficiency was measured as the average time-to-review per PR, capturing the complete duration from input to output generation.  
Human reviewers were timed manually, whereas response times for LMs and SonarQube were extracted from system log files.

\section{Results}
\subsection{Results of Code Change Quality Estimation}
\begin{figure}[!htbp]
  \centering
  \includegraphics[width=\linewidth]{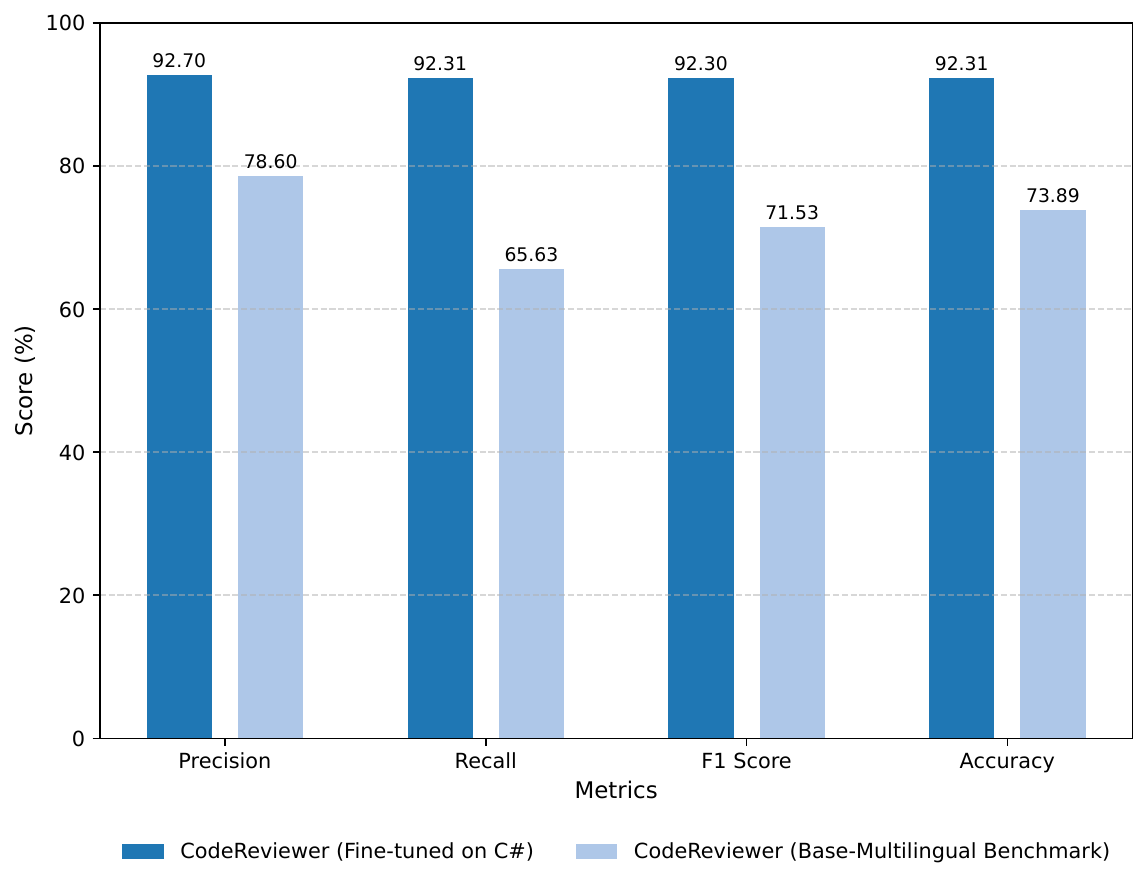}
  \caption{Code change quality estimation performance of CodeReviewer fine-tuned on monolingual C\# dataset compared to its base multilingual version.}
  \label{fig:qualityEstimationPerformance}
  \Description{Bar chart comparing CodeReviewer’s precision, recall, F1, and accuracy for monolingual and multilingual training settings on the code change quality estimation task.}
\end{figure}

Figure~\ref{fig:qualityEstimationPerformance} shows the performance of CodeReviewer on the code change quality estimation task, comparing the monolingual C\# fine-tuned variant with the multilingual baseline by \citet{li2022automating}. Despite being trained on a smaller dataset, the monolingual version outperformed the multilingual model across all evaluation metrics, including precision, recall, F1 score, and accuracy. This consistent improvement suggests that domain and language specific fine-tuning can effectively enhance the model's ability to identify relevant code changes, even with limited training data.
\subsection{Results of Review Comment Generation}
\label{sec:results-cg}
Table~\ref{tab:cg-results} provides a comparative evaluation of the models on the review comment generation task, using both automated (BLEU-4) and human-centered (Information and Relevance) metrics. A key finding is that no single model dominates across all evaluation criteria, indicating trade-offs between lexical similarity, informativeness, and contextual relevance.

CodeReviewer, when fine-tuned on monolingual C\# with English-only comments, achieves the highest BLEU-4 score, reflecting strong alignment with reference phrasing. However, it slightly lags behind the base model in Information and Relevance. Its bilingual variant performs similarly in BLEU-4 but worse in human evaluations, suggesting that linguistic inconsistency in training data may reduce fluency and coherence. The base CodeReviewer—pre-trained specifically for code review on a multilingual corpus—achieves lower BLEU but competitive human-rated scores, highlighting a trade-off between lexical similarity and perceived quality.

CodeLlama-7B, fine-tuned on the same monolingual setting, shows balanced performance. While its BLEU-4 score is slightly lower than CodeReviewer’s, it achieves the highest Information and strong Relevance scores, demonstrating that instruction-tuned, code-pretrained LLMs can generate informative and context-aware comments after fine-tuning. Its base model, trained on multilingual data, performs considerably worse across all metrics, underscoring the importance of task adaptation.

DeepSeek-R1-Distill-Llama-8B presents a different trade-off. Fine-tuning on monolingual data significantly boosts BLEU-4, but slightly reduces Information and Relevance compared to its base model. Notably, the base model achieves the highest Relevance score overall, despite its low BLEU-4, suggesting that fine-tuning may improve lexical fidelity at the expense of general reasoning and feedback richness.

\begin{table*}[!htbp]
  \centering
  \begin{threeparttable}
    \caption{Comparison of models on the review comment generation task, categorized by PL scope and NL of comments.}
    \label{tab:cg-results}
    \begin{tabular}{lccccc}
      \toprule
      \textbf{Model} & \textbf{PL Scope} & \textbf{NL of Comments} & \textbf{BLEU-4} & \textbf{Information} & \textbf{Relevance} \\
      \midrule
      CodeReviewer fine-tuned on \texttt{C\#} (English+German) & Mono-PL & English+German & 8.76 & 1.20 & 1.41 \\
      CodeReviewer fine-tuned on \texttt{C\#} (English) & Mono-PL & English-only & \textbf{9.08} & 3.47 & 3.16 \\
      CodeLlama-7B fine-tuned on \texttt{C\#} & Mono-PL & English-only & 8.08 & \textbf{3.80} & 3.67 \\
      DeepSeek-R1-Distill-Llama-8B fine-tuned on \texttt{C\#} & Mono-PL & English-only & 8.12 & 3.48 & 3.61 \\
      \midrule
      CodeReviewer base \cite{li2022automating} & Multi-PL & Multilingual & 5.32 & 3.60 & 3.20 \\
      CodeLlama-7B base \cite{haider2024prompting} & Multi-PL & Multilingual & 5.58 & 3.13 & 3.45 \\
      DeepSeek-R1-Distill-Llama-8B base \cite{guo2025deepseek} & Multi-PL & Multilingual & 3.45 & 3.61 & \textbf{3.93} \\
      \bottomrule
    \end{tabular}
    \begin{tablenotes}
      \small
      \item The best score for each metric is shown in bold.
    \end{tablenotes}
  \end{threeparttable}
\end{table*}
\subsection{Results of Code Refinement}
\begin{figure}[!htbp]
  \centering
  \includegraphics[width=\linewidth]{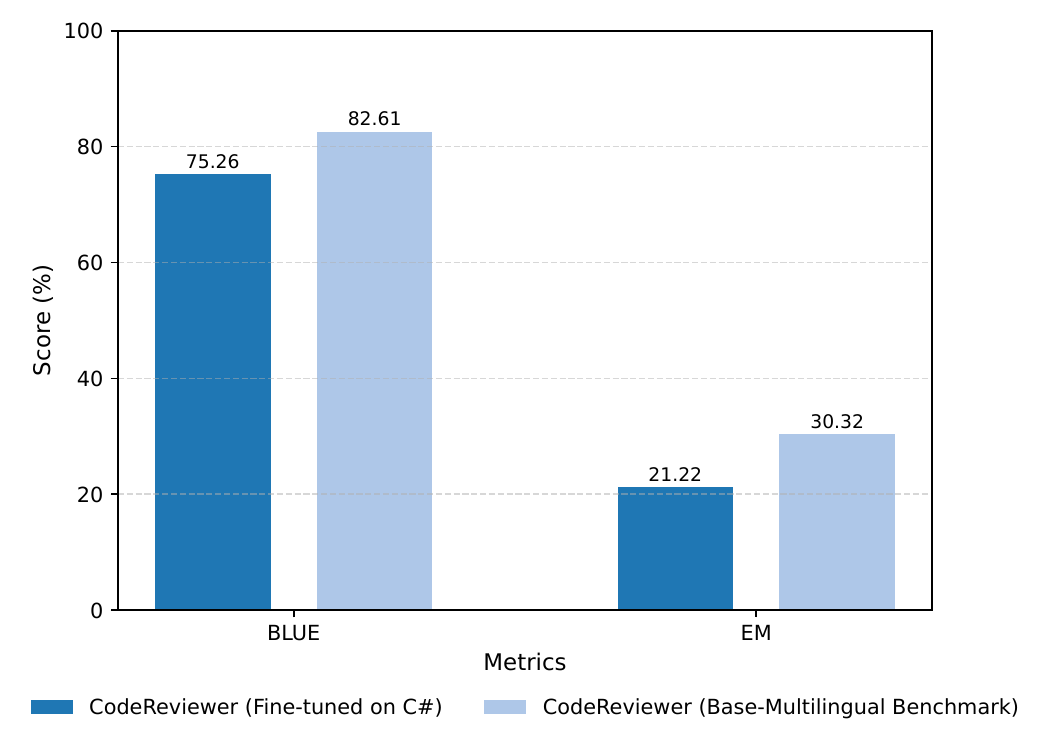}
  \Description{Bar chart showing performance comparison of CodeReviewer on code refinement task.}
  \caption{Code refinement performance of CodeReviewer fine-tuned on monolingual C\# dataset compared to its base multilingual version.}
  \label{fig:codeRefinment-data}
\end{figure}

Figure~\ref{fig:codeRefinment-data} shows the results of the code refinement task, comparing the performance of CodeReviewer when fine-tuned on a monolingual C\# dataset versus its original multilingual version from the CodeReviewer benchmark~\cite{li2022automating}. Unlike the other tasks, fine-tuning on monolingual data led to a decline in both BLEU and EM scores. This suggests that domain and language specific fine-tuning, while effective for tasks closely tied to phrasing or review context, may be less suitable for tasks requiring a broader generalization or exposure to diverse examples. One likely reason is the significantly smaller and less varied size of the monolingual fine-tuning dataset, which may have limited the model’s ability to learn complex correction patterns compared to the multilingual training corpus provided by \citet{li2022automating}.

\subsection{Comparison of LMs, SonarQube, and Human Reviewers}
\label{sec:results-human-sonarqube}
\begin{figure}[tbp]
  \centering
  \includegraphics[width=\linewidth]{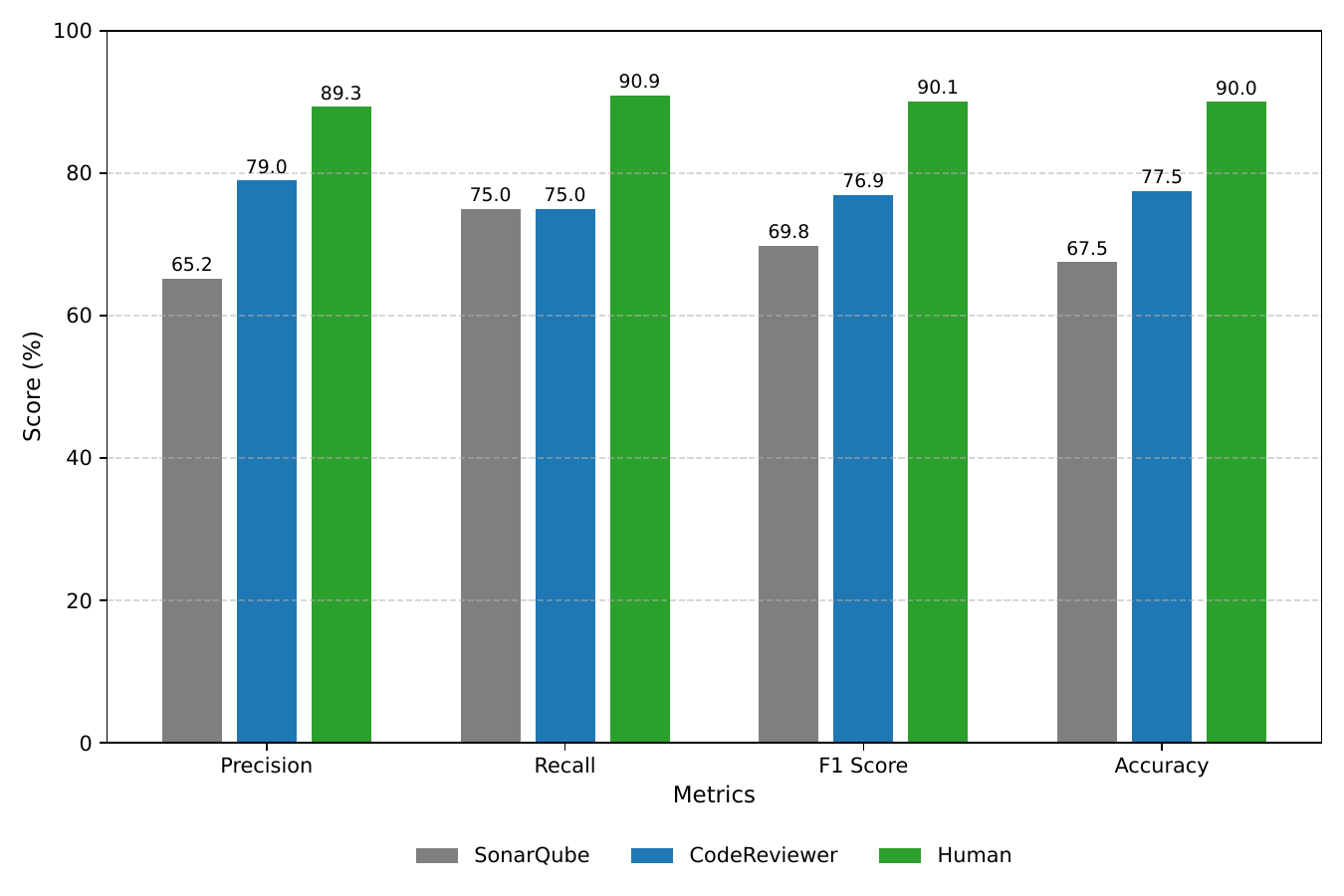}
  \Description{Comparison of methods on the code change quality estimation task.}
  \caption{Comparison of SonarQube, CodeReviewer, and human reviewers on 40 PRs in the \textit{Code Change Quality Estimation} task. The CodeReviewer was fine-tuned on C\# with English comments.}
  \label{fig:compare-QE}
\end{figure}

\begin{figure}[tbp]
  \centering
  \includegraphics[width=\linewidth]{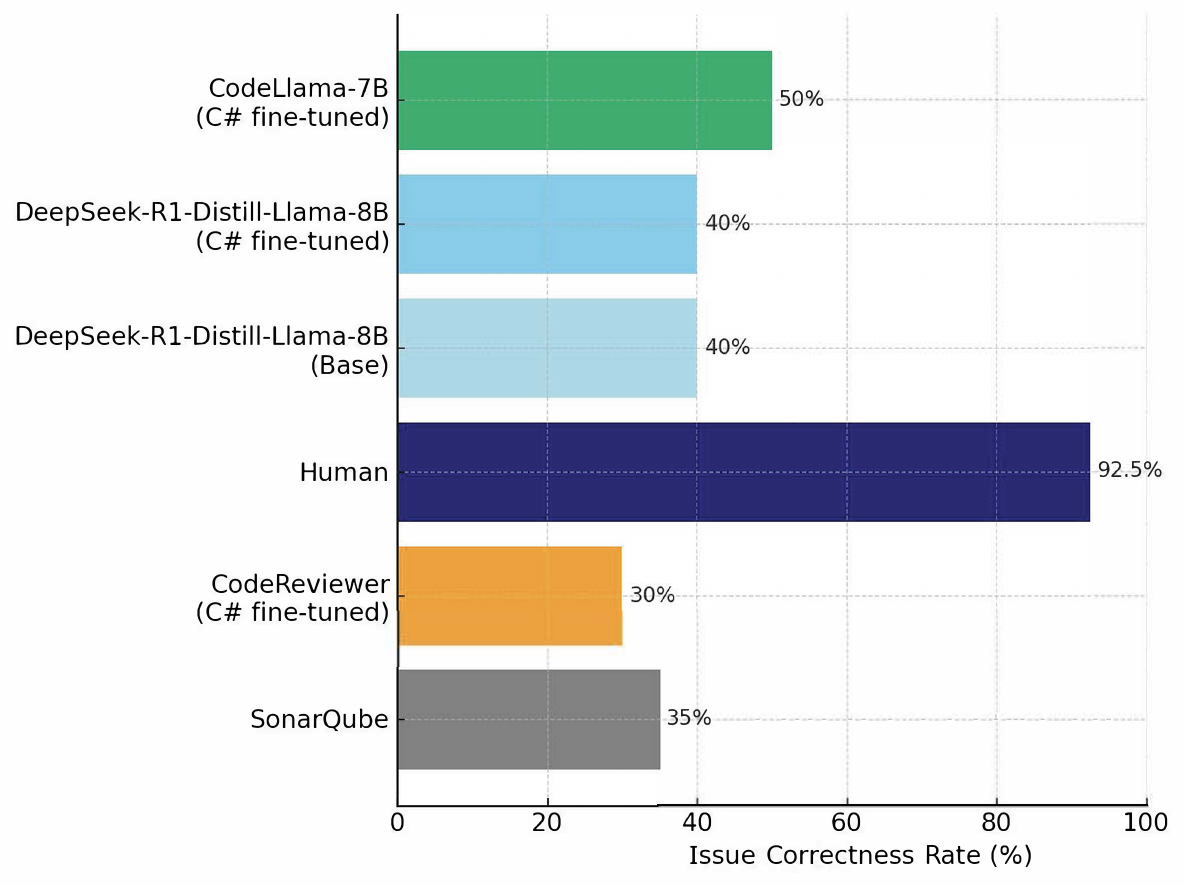}
  \Description{Comparison of methods on the review comment generation task.}
  \caption{Comparison of SonarQube, CodeReviewer, and human reviewers on 40 PRs in the \textit{Review Comment Generation} task.  
  The CodeReviewer was fine-tuned on C\# with English comments.  
  The \textit{Issue Correctness Rate} represents the proportion of generated comments judged as “correct,” i.e., accurately addressing the corresponding code issue according to the ground truth.}
  \label{fig:compare-CG}
\end{figure}

Figures~\ref{fig:compare-QE} and~\ref{fig:compare-CG} present the comparative performance of human reviewers, SonarQube, and LMs across the two main evaluation tasks—code change quality estimation and review comment generation—based on a subset of 40 PRs from our industrial C\# dataset.

\textit{Code Change Quality Estimation.} For the binary classification task of identifying whether a code change requires a review comment, both SonarQube and the fine-tuned CodeReviewer performed close to human reviewers in terms of accuracy, precision, recall, and F1 score (see Figure~\ref{fig:compare-QE}). A quantitative analysis of prediction errors across issue categories revealed notable performance differences between the methods. SonarQube showed strong performance in detecting rule-based issues, such as security vulnerabilities and maintainability concerns. However, it struggled with more context-dependent categories like performance optimizations and logic errors. In contrast, the fine-tuned CodeReviewer delivered more balanced results across all issue types, suggesting better generalization capabilities for identifying both rule-based and context-sensitive quality issues.

\textit{Review Comment Generation.} In the review comment generation task, the performance gap between human reviewers and automated methods became more pronounced (see Figure~\ref{fig:compare-CG}). 
Human reviewers consistently produced comments that correctly addressed the underlying code issues, resulting in the highest \textit{Issue Correctness Rate}.  
Among the language models, CodeLlama-7B (fine-tuned on C\#) achieved the best alignment with the ground truth, followed closely by CodeReviewer and DeepSeek-R1-Distill-Llama-8B.  
Notably, the DeepSeek model produced almost identical comments before and after fine-tuning, suggesting that the fine-tuning process reduced its reasoning variety while improving lexical precision.  
Overall, fine-tuned models narrowed—but did not eliminate—the gap to human reviewers in terms of correctness and issue coverage.

\textit{Efficiency Trade-offs.} Table~\ref{tab:review_time} presents the average review time per PR for each method. Human reviewers typically required over five minutes per PR, depending on the complexity of the changes. SonarQube took between three to five minutes due to its thorough static analysis. In contrast, LMs significantly reduced review time, generally completing reviews in under one minute. Among them, the DeepSeek-R1-Distill-Llama-8B base was the slowest, likely due to its more elaborate reasoning-based responses.

\begin{table}[!htbp]
\centering
\small
\caption{Average time-to-review per PR by method. Measurements reflect end-to-end processing time across the full evaluation dataset, not limited to the 40 PRs used for human comparison.}
\label{tab:review_time}
\begin{tabular}{@{}lc@{}}
\toprule
\textbf{Method} & \textbf{Time-to-Review (min)} \\
\midrule
DeepSeek-R1-Distill-Llama-8B base & 1--3 \\
Other LMs & $<$1 \\
SonarQube & 3--5 \\
Human Reviewers & 5--7 \\
\bottomrule
\end{tabular}
\end{table}

\section{Discussion}
\subsection{Effectiveness of Monolingual Fine-Tuning for CodeReviewer on Code Change Estimation}
Our findings provide empirical evidence that monolingual fine-tuning on C\# can enhance the performance of CodeReviewer, particularly in the Code Change Quality Estimation task. Despite being trained on a smaller dataset, the monolingual fine-tuned variant consistently outperformed its multilingual counterpart. This supports previous research indicating that language-specific adaptation helps models better capture syntactic patterns and task-specific signals \cite{ahmad2021unified, chen2022transferability}.

An important methodological factor contributing to this outcome was our negative sample selection strategy. We deliberately included only the final code hunk of each PR as negative examples (i.e., those not requiring a review comment). These hunks typically contained fewer and more stable changes, making the learning task less noisy. This targeted sampling likely helped the model to more effectively learn decision boundaries between review-worthy and non-review-worthy code changes, echoing findings from prior work on label quality and class balance in code classification tasks \cite{hellendoorn2021towards}.

Overall, these findings demonstrate that domain- and language-specific fine-tuning is particularly beneficial for structured classification-oriented review tasks such as determining whether a code change requires human intervention.

\subsection{Influence of Model Type, PL/NL Scope, and Fine-Tuning Design on Review Comment Generation}
Our results highlight that while monolingual fine-tuning on C\# improves LM performance in Review Comment Generation, the relationship between BLEU-4 scores and human-perceived comment quality is not straightforward. Although the fine-tuned CodeReviewer achieved the highest BLEU-4 score, human evaluators consistently rated the fine-tuned CodeLlama-7B higher in both Information and Relevance. This discrepancy underscores known limitations of BLEU-4 as a sole quality indicator for code-related NL generation tasks \cite{li2022automating, zhao2023survey}.

One likely reason for CodeReviewer’s high BLEU-4 score lies in its learned use of token patterns such as emojis, which appeared frequently in the training data. Although such token patterns increased lexical similarity and improved BLEU-4 scores, they did not necessarily contribute to the informativeness or contextual relevance of the generated comments. Moreover, the mixed-language composition (English and German) of the fine-tuning data for some CodeReviewer variants likely introduced linguistic inconsistencies, occasionally resulting in comments that were difficult to understand or exhibited unnatural language mixing. This finding aligns with studies showing that inconsistencies in training data language composition can negatively affect LMs generation quality \cite{cihan2024automated}.

A noteworthy observation emerged with DeepSeek-R1-Distill-Llama-8B. Before fine-tuning, this reasoning-focused LLM produced comments that human evaluators perceived as more context-aware and insightful. However, after fine-tuning on our instruction-light dataset, the model’s BLEU-4 score improved, but its ability to generate reasoning-rich, explanatory comments declined. This suggests that for models like DeepSeek-R1, maintaining reasoning capabilities may require specialized fine-tuning approaches that preserve or enhance CoT reasoning during adaptation.

Overall, these findings demonstrate that while monolingual fine-tuning helps align LMs with the target PL and improves surface-level lexical similarity, it may introduce trade-offs in reasoning quality and linguistic richness. These trade-offs are further shaped by the underlying pretraining paradigms of the LMs. Their divergent behaviors after fine-tuning illustrate how pretraining objectives influence adaptability to domain-specific tasks like code review. We recommend integrating more linguistically balanced and instruction-rich datasets—possibly including human-annotated reasoning traces—to better support both fluency and depth in generated review comments.

\subsection{Effectiveness of Monolingual Fine-Tuning for CodeReviewer on Code Refinement}
However, these benefits did not extend to the Code Refinement task. The monolingual fine-tuned model underperformed the multilingual baseline in both BLEU and exact match scores. One plausible reason is the limited size and diversity of the monolingual training data, which may have constrained the model’s ability to generalize to broader or more complex edit patterns. Prior work has emphasized that code generation tasks, such as refinement, benefit from large and diverse datasets \cite{li2022automating, lin2023cct5}. Additionally, translation-related inconsistencies in the dataset may have introduced noise, a known threat to generation quality in multilingual settings \cite{zhao2023survey}.

In summary, these observations indicate that fine-tuning is not uniformly beneficial across all ACR stages. While it strengthens discriminative capabilities, it can limit generative diversity and robustness in more complex code transformation tasks.

\subsection{Comparison of Fine-Tuned LMs, SonarQube, and Human Reviewers}
Our findings reveal that while monolingual fine-tuning improves LLM performance across code review tasks, human reviewers still consistently outperform both LLM-based models and ASATs like SonarQube. This performance gap was especially pronounced in the Review Comment Generation task, where human reviewers achieved substantially higher semantic accuracy than all automated methods. These results align with prior studies showing that human reviewers provide more actionable, context-aware, and nuanced feedback than current automated tools \cite{hellendoorn2021towards, cihan2024automated}.

In the Code Change Quality Estimation task, the fine-tuned CodeReviewer showed a noticeable drop in performance when evaluated on our diverse, industrial PR sample compared to its controlled test set results. This degradation likely reflects the domain shift and code diversity in real-world PRs—an issue commonly noted in prior LLM-based software engineering studies \cite{li2022automating, lin2024}. Unlike more homogeneous training and testing datasets, our industrial PRs represented varied coding styles, standards, and project-specific conventions, reinforcing concerns raised by \citet{vassallo2020developers} regarding tool generalizability.

SonarQube, as expected from prior ASAT studies \cite{beller2016analyzing, charoenwet2024empirical}, performed well in identifying rule-based issues like code smells and security vulnerabilities. However, its lack of semantic understanding limited its ability to detect deeper logic-related or architectural flaws—consistent with earlier critiques of ASATs\cite{panichella2015would}.

The LLM-based models, particularly the fine-tuned CodeLlama-7B, demonstrated better generalization across issue categories, especially for refactoring suggestions and clarity improvements. This aligns with recent findings that LLMs can capture higher-level code semantics more effectively than rule-based tools \cite{lu2023llama}. However, even the best-performing LLM still lagged behind human reviewers in delivering comprehensive, context-sensitive feedback—a limitation similarly reported in \cite{cihan2024automated}.

An important observation was the trade-off between review speed and review quality. While both LMs and SonarQube completed reviews significantly faster than humans (often under one minute), their feedback lacked the depth and accuracy required for critical code assessment. This reflects a broader trend in AI-assisted code review research, where efficiency gains often come at the expense of review depth and trustworthiness \cite{zhao2023survey}.

Overall, while AI and SonarQube offer substantial efficiency gains, they still fall short of human reviewers in producing high-quality, context-aware feedback. These results suggest that AI tools can serve as valuable assistants to accelerate the review process but cannot yet fully replace expert human judgment in critical code assessment tasks. A hybrid approach—combining the speed of AI with human expertise—may represent the most effective strategy for industrial code review workflows.

\section{Implications}
Our findings offer practical implications for both industry practitioners and researchers.

\textbf{For practitioners}, fine-tuned LMs can serve as fast and reasonably accurate assistants in CI pipelines. While they do not match human-level performance—particularly in nuanced or complex review scenarios—their speed and early-issue detection capabilities make them valuable for triage and prioritization tasks. In particular, LMs can help filter routine PRs, reducing the cognitive load on human reviewers.

Integrating LM-based tools with static analyzers like SonarQube can lead to more balanced workflows. Whereas SonarQube is effective in flagging rule-based issues (e.g., security or maintainability violations), LMs offer more linguistically rich and context-sensitive feedback. This division of labor enables efficient issue coverage across both syntactic and semantic dimensions.

Another benefit is cost-efficiency. The evaluated LMs are open-source and license-free, providing accessible solutions for organizations operating under budget constraints. Moreover, task-specific monolingual fine-tuning emerged as a practical strategy to improve model effectiveness in single-language environments like C\#.

\textbf{For researchers}, our results raise concerns about relying solely on automated metrics such as BLEU to assess code review quality. Human-centered evaluations remain crucial for capturing relevance and informativeness. In addition, the performance drop observed in reasoning-intensive models like DeepSeek-R1-Distill after fine-tuning suggests a need for strategies that preserve reasoning capabilities—potentially via chain-of-thought data or multi-objective optimization approaches.

\section{Threats to Validity}
Our study has several validity threats, discussed in terms of internal, external, and construct validity.

\textbf{Internal Validity.} One threat is the limited size and language scope of the monolingual C\# dataset. The small volume of high-quality code-review pairs may have restricted the models' exposure to diverse code patterns, especially in the Code Refinement task. Additionally, translating German review comments into English may have introduced semantic noise, affecting both model learning and human evaluation.
Although translation quality was manually validated with high adequacy, fluency, and inter-annotator agreement, minor translation noise may still persist and could have subtly influenced model behavior.

Evaluation variability presents another threat. Although six professional software engineers with varying experience levels rated the outputs, subjective bias remains possible. 
We mitigated this by averaging scores and resolving annotation disagreements through consensus-based discussion. 
To further minimize potential evaluator bias, all raters were independent of the dataset construction and translation validation processes, and human evaluation tasks (manual reviewing vs. model-output assessment) were conducted in separate sessions. 
Despite these precautions, residual subjective variability cannot be entirely eliminated and represents a remaining internal validity limitation.

\textbf{External Validity.} Our experiments focused on C\# from a single industrial partner, supplemented with open-source data. This limits generalizability to other languages or organizational contexts. Replications on other languages and datasets are needed to confirm broader applicability.

\textbf{Construct Validity.} Our choice to employ BLEU metrics follows the same evaluation protocol as the original \textit{CodeReviewer} study~\cite{li2022automating}, ensuring direct comparability with their reported C\# results. 
Although more specialized metrics such as CodeBLEU~\cite{ren2020codebleu} or CrystalBLEU~\cite{CrystalBLEU2023} offer better code-sensitive evaluation, applying them solely to our models—without equivalent outputs from Microsoft’s unpublished checkpoints—would have compromised cross-study comparability. 
Relying on BLEU-based metrics is a known limitation, as these capture lexical similarity but may overlook semantic or structural correctness. 
To mitigate this, we complemented BLEU-4 (used for comment generation) and BLEU (used for code refinement) with human-centered Information and Relevance ratings. Still, evaluating LLM-generated review quality remains challenging. 

Another construct-related limitation concerns the alignment between refinement outputs and functional correctness. 
While all generated code snippets were syntactically valid and successfully parsed by the compiler, execution-level or unit-test validation could not be performed due to the absence of complete build contexts in the industrial repositories. 
As a result, BLEU and EM scores capture lexical and structural similarity but do not fully account for functional or semantic correctness of the generated code.

The choice of baselines may influence results. The choice of baselines may influence results. We used strong open-source models, which represent current state-of-the-art (SOTA) approaches specifically developed for ACR. 
In contrast, general-purpose frontier LLMs such as GPT-4 or GPT-5 are not explicitly optimized for ACR and cannot be fine-tuned or benchmarked under the same controlled conditions. Therefore, our focus remained on models that are technically comparable, reproducible, and directly aligned with the ACR task objectives. Additionally, comparisons with general-purpose frontier LLMs were beyond our budget constraints. For practical reasons, we did not perform multi-run evaluations for each configuration, meaning that variance across runs remains possible for non-deterministic models like DeepSeek-R1-Distill-Llama-8B.

Also, for practical reasons, we did not perform multi-run evaluations for each setting, meaning variance across runs remains possible for non-deterministic models like DeepSeek-R1-Distill-Llama-8B. 

\section{Conclusion and Future Work}
This study examined whether fine-tuning different types of LMs, including a PLM specialized in code review (CodeReviewer), a LLM pretrained on code (CodeLlama-7B), and a general-purpose LLM (DeepSeek-R1-Distill), on monolingual C\# data that combines public benchmarks with proprietary industrial code yields performance gains over their original multilingual versions. We evaluated these models across three core code review tasks and compared their performance against both an ASAT and human reviewers.

The overall findings show that human reviewers still deliver the highest quality and most context aware feedback. They are able to capture nuances and understand the deeper implications of code changes in ways that current LMs and ASATs cannot. Nevertheless, the AI-based approaches and SonarQube offer valuable benefits, such as significantly faster review times and consistent output. For instance, the AI models performed well in quickly identifying whether a PR needed a review, even if their generated comments were not always as detailed or accurate as those provided by human experts.
These findings highlight the potential of integrating AI-driven code review tools into existing workflows. By combining the speed and efficiency of automated systems with the deep, contextual understanding of human reviewers, it is possible to create a more balanced and effective approach to maintaining high software quality. More detailed information on our experimental setups, parameter settings, and evaluation procedures can be found in our repository.

Future research may expand to other object-oriented languages to assess cross-language generalizability. Incorporating structural representations like Abstract Syntax Trees could enhance model understanding. Additionally, applying CoT fine-tuning may improve the reasoning transparency of LLM-generated feedback. A systematic analysis of hyperparameter sensitivity could further strengthen the robustness and interpretability of future fine-tuning efforts. Future studies could extend this work by incorporating controlled API-based evaluations of general-purpose SOTA LLMs like GPT-5 to examine whether their broader reasoning and linguistic capacities translate into higher review fidelity.
Such experiments would help bridge the gap between domain-specific ACR fine-tuning and general-purpose LLM performance, offering a more complete picture of the trade-offs between specialization and generalization in automated code review.

Future studies could also incorporate statistical significance testing across multiple fine-tuning runs and hyperparameter configurations to quantify the robustness of performance differences. 
Such analyses would help determine whether observed variations stem from model behavior or stochastic effects during optimization.
\bibliographystyle{ACM-Reference-Format}
\bibliography{bibliography}

\end{document}